# Energy of One-Dimensional Diatomic Elastic Granular Gas:
# Theory and Molecular Dynamics Simulation

Siti Nurul Khotimah[a]*, Sparisoma Viridi[a], Widayani[a] and Abdul Waris[a]

[a]Nuclear Physics and Biophysics Research Division
Faculty of Mathematics and Natural Sciences
Institut Teknologi Bandung, Indonesia
*Email nurul@fi.itb.ac.id

**Abstract**

One-dimensional ideal diatomic gas is simulated through possible types of motion of a molecule. Energy of each type of its motion is calculated from theory and numerical method. Calculation of kinetic energy of an atom in translational-vibrational motion is not analytically simple, but it can be solved by numerical method of molecular dynamic simulation. This paper justifies that kinetic energy of a diatomic molecule can be determined by two different approaches. The first is the sum of kinetic energy of each atom and second is the sum of kinetic energy of translational motion and vibrational motion.

**Keywords**: one-dimension, granular gas, diatomic molecule, simulation

## Introduction

Motion of diatomic gas molecules, which is temperature dependent, is contributed from three types of motion: translational, rotational, and vibrational. It is already obvious to consider that those types of motion are independent, which lead to the well known concept of equipartition energy and degree of freedom. This concept has been theoretically established [1, 2]. Students generally do not realize that a diatomic molecule may make a motion contains more than one type of motion, as we have been observed recently. Therefore, it is required to build a model of motion of a diatomic molecule in order to connect between that abstract concept and students' concrete understanding.

One way to model the motion of gas molecules is to use a simulated granular material. An atom is modeled as a ball. One method that is commonly used is the soft-sphere [3] using molecular dynamics method implemented Gear predictor-corrector algorithm [4]. As a first step, the problem is limited to one-dimensional case. There is a study on collision properties [5] and energy transport in [6], free cooling [7], and kinetic description [8] of one-dimensional granular gas. One-dimensional ideal diatomic gas which is simulated through motion of a molecule in order to observe its possible types of motion is reported in this work. Energy calculation from theory and numerical method is also discussed.

## Theoretical Background

A diatomic gas molecule is modeled as two atoms which are connected by an ideal spring. For one-dimensional motion, the molecule is moving with its trajectories are straight lines. Molecule can move translational, vibrational, or translational-vibrational. The energy of this molecule is discussed by reviewing the kinetic energy and potential energy separately.

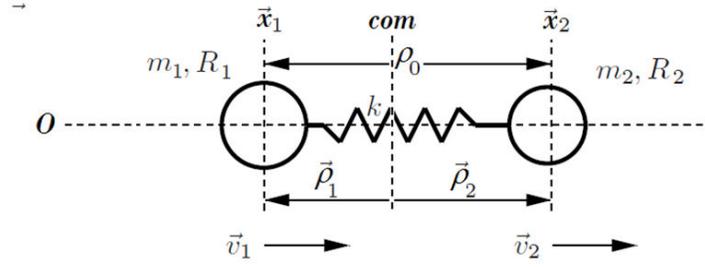

Figure 1. A model of a diatomic molecule as two atoms connected by an ideal spring.

Mass, position, velocity, momentum and kinetic energy of the $i$th atom are denoted by $m_i, \vec{x}_i, \vec{v}_i, \vec{p}_i$ and $K_i$. Index $s$ denotes for system of a diatomic molecule. Kinetic energy of the system is

$$K_s = K_1 + K_2 = \tfrac{1}{2} m_1 v_1^2 + \tfrac{1}{2} m_2 v_2^2 = \frac{p_1^2}{2m_1} + \frac{p_2^2}{2m_2} \tag{1}$$

If a diatomic molecule is treated as a point, coordinate and momentum of this point correspond to the center of mass (com) of the molecule so that there are two groups of variables. The first is related to the center of mass, while the latter to internal system denoted by Greek symbol. Kinetic energy of the system is then written as [3]:

$$K_s = \frac{p_s^2}{2m_s} + \frac{1}{2} \mu \, \dot{\rho}_{12}^2 \tag{2}$$

where $\mu$ is the reduced mass of the system

$$\mu = \frac{m_1 m_2}{m_1 + m_2} \tag{3}$$

$\rho_{12}$ is distance between the two atoms and $\dot{\rho}_{12}$ is its time derivative.

$$\vec{\rho}_{12} = \vec{\rho}_1 - \vec{\rho}_2 = \vec{x}_1 - \vec{x}_2 = \vec{x}_{12} \tag{4}$$

If the spring between two atoms in the system has spring constant $k$ then potential energy of the system is

$$U = \frac{1}{2} k (\rho_0 - \rho_{12})^2 \tag{5}$$

where $\rho_0$ is an equilibrium distance.

The energy of the molecule is

$$E = K_s + U = \left( \tfrac{1}{2} m_1 v_1^2 + \tfrac{1}{2} m_2 v_2^2 \right) + \frac{1}{2} k (\rho_0 - \rho_{12})^2 \tag{6}$$

or

$$E = E_{trans} + E_{vib} = \left( \frac{p_s^2}{2m_s} \right) + \left[ \frac{1}{2} \mu \, \dot{\rho}_{12}^2 + \frac{1}{2} k (\rho_0 - \rho_{12})^2 \right] \tag{7}$$

**Simulation Procedure**

Two grains which are connected with an ideal spring are used to simulate a diatomic molecule. The spring force $\vec{S}_{ij}$ between the two grains is:

$$\vec{S}_{ij} = k (\rho_0 - \rho_{ij}) \hat{\rho}_{ij} - \gamma \, \vec{v}_{ij} \tag{8}$$

Period of vibrational motion is

$$\tau_{vib} = 2\pi \sqrt{\frac{\mu}{k}} \tag{9}$$

Gear predictor-corrector algorithm of fifth order [4] is chosen in the molecular dynamics method used in the simulation, which has two steps: prediction step (written with upper index $p$) and correction step for every particular grain. The first step is formulated as

$$\begin{pmatrix} \vec{r_0}^{\,p}(t+\Delta t) \\ \vec{r_1}^{\,p}(t+\Delta t) \\ \vec{r_2}^{\,p}(t+\Delta t) \\ \vec{r_3}^{\,p}(t+\Delta t) \\ \vec{r_4}^{\,p}(t+\Delta t) \\ \vec{r_5}^{\,p}(t+\Delta t) \end{pmatrix} = \begin{pmatrix} 1 & 1 & 1 & 1 & 1 & 1 \\ 0 & 1 & 2 & 3 & 4 & 5 \\ 0 & 0 & 1 & 3 & 6 & 10 \\ 0 & 0 & 0 & 1 & 4 & 10 \\ 0 & 0 & 0 & 0 & 1 & 5 \\ 0 & 0 & 0 & 0 & 0 & 1 \end{pmatrix} \begin{pmatrix} \vec{r_0}(t) \\ \vec{r_1}(t) \\ \vec{r_2}(t) \\ \vec{r_3}(t) \\ \vec{r_4}(t) \\ \vec{r_5}(t) \end{pmatrix}. \tag{10}$$

And the correction step will give the corrected value of $\vec{r_n}(t+\Delta t)$ through

$$\begin{pmatrix} \vec{r_0}(t+\Delta t) \\ \vec{r_1}(t+\Delta t) \\ \vec{r_2}(t+\Delta t) \\ \vec{r_3}(t+\Delta t) \\ \vec{r_4}(t+\Delta t) \\ \vec{r_5}(t+\Delta t) \end{pmatrix} = \begin{pmatrix} \vec{r_0}^{\,p}(t+\Delta t) \\ \vec{r_1}^{\,p}(t+\Delta t) \\ \vec{r_2}^{\,p}(t+\Delta t) \\ \vec{r_3}^{\,p}(t+\Delta t) \\ \vec{r_4}^{\,p}(t+\Delta t) \\ \vec{r_5}^{\,p}(t+\Delta t) \end{pmatrix} + \begin{pmatrix} c_0 \\ c_1 \\ c_2 \\ c_3 \\ c_4 \\ c_5 \end{pmatrix} \Delta\vec{r_2}(t+\Delta t), \tag{11}$$

with

$$\Delta\vec{r_2}(t+\Delta t) = \vec{r_2}(t+\Delta t) - \vec{r_2}^{\,p}(t+\Delta t) \tag{12}$$

The term $\vec{r_n}(t+\Delta t)$ is defined as

$$\vec{r_n}(t) = \frac{(\Delta t)^n}{n!} \left[ \frac{d^n \vec{r_0}(t)}{dt^n} \right], \tag{13}$$

where $\vec{r_0}$ is position of a grain. The term $\vec{r_2}(t+\Delta t)$ in correction term in Equation (12) is obtained from Newton' second law of motion. For example, particle $i$ has

$$\left[ \vec{r_2}(t+\Delta t) \right]_i = \frac{(\Delta t)^2}{m_i} \sum_{j \neq i} \vec{S}_{ij}(t+\Delta t). \tag{14}$$

The left part of Equation (11) is calculated using $\vec{r_n}^{\,p}(t+\Delta t)$.

**RESULTS AND DISCUSSION**
**A. Kinetic, Potential, and Mechanical Energies**
A simulation result for a diatomic molecule in translational, vibrational, and translational-vibrational motion are given in Figure 2. Position of the two atoms as a function of time shown for each case. Simulation parameters are $m_1 = m_2 = 2$, $k = 100$, $\gamma = 0$, $R_1 = R_2 = 0.01$. The two atoms are initially at $\bar{x}_1(t=0) = -0.5$ and $\bar{x}_2(t=0) = 0.5$ so that its equilibrium distance $\rho_0 = 1.0$.

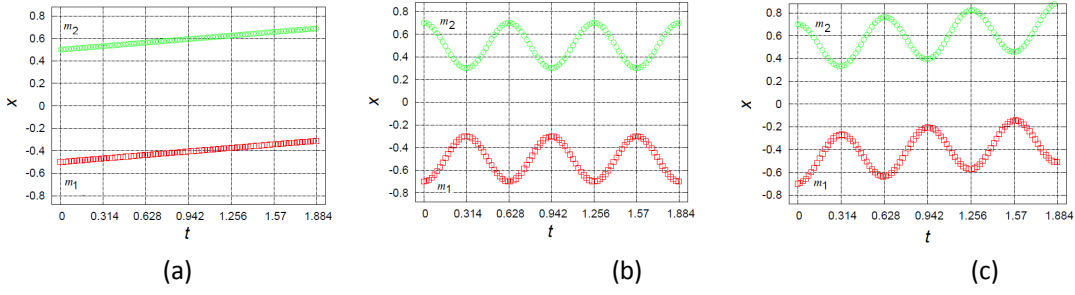

(a)                              (b)                              (c)

Figure 2. Position-time graph for the two atoms of a molecule:
(a) translational, (b) vibrational, and
(c) translational-vibrational motion.

The translational velocity of the atoms in Figure 2(a) is $\vec{v}_1 = \vec{v}_2 = 0.1$ so that the distance between the atoms is constant ($\rho_{12} = \rho_0 = 1.0$) and its time derivative is zero ($\dot{\vec{\rho}}_{12} = \vec{v}_{12} = 0$). Kinetic, potential, and total energy for this translational motion are

$$K_{trans} = \tfrac{1}{2} m_1 v_1^2 + \tfrac{1}{2} m_2 v_2^2 = 0.01 + 0.01 = 0.02 \tag{15}$$

$$U_{trans} = \frac{1}{2} k (\rho_0 - \rho_{12})^2 = 0 \tag{16}$$

$$E_{trans} = K_{trans} + U_{trans} = 0.02 \tag{17}$$

In Figure 2(b) each atom is deflected outward 0.2 from their initial positions so that molecule vibrates with amplitude A = 0.4 at period $\tau_{vib}$= 0.2π. The distance between the two atoms from its equilibrium changes with time in sinusoidal function.

$$\left(\rho_{ij} - \rho_0\right) = A\cos(\omega_{vib} t)$$

Its relative velocity between the two atoms is also in sinusoidal function.

$$\dot{\rho}_{ij} = -A\omega_{vib}\sin(\omega_{vib} t)$$

For this vibrational motion, position of the center of mass is fixed ($\vec{v}_s = 0$ and $\vec{p}_s = 0$). Kinetic, potential, and total energy for this vibrational motion are

$$K_{vib} = \frac{1}{2}\mu(\dot{\rho}_{12})^2 = \frac{1}{2}\mu\left(-A\omega_{vib}\sin(\omega_{vib}t)\right)^2 = \frac{1}{2}kA^2\sin^2(\omega t) = 8\sin^2(10t) \tag{18}$$

$$U_{vib} = \frac{1}{2}k(\rho_{12} - \rho_0)^2 = \frac{1}{2}kA^2\cos^2(\omega_{vib}t) = 8\cos^2(10t) \tag{19}$$

$$E_{vib} = K_{vib} + U_{vib} = \frac{1}{2}kA^2 = 8 \tag{20}$$

Figure 2(c) shows a translational and vibrational motion. Kinetic, potential, and total energy for this motion are

$$U = \frac{1}{2}k(\rho_{12} - \rho_0)^2 = \frac{1}{2}kA^2\cos^2(\omega_{vib}t) = 8\cos^2(10t) \tag{21}$$

$$K = \frac{p_s^2}{2m_s} + \frac{1}{2}\mu(\dot{\rho}_{12})^2 = \frac{1}{2}m_s v_s^2 + \frac{1}{2}\mu\left(-A\omega_{vib}\sin(\omega_{vib}t)\right)^2$$

$$= 0.02 + \frac{1}{2}kA^2\sin^2(\omega_{vib}t) = 0.02 + 8\sin^2(10t) \tag{22}$$

$$E = K + U = 8.02 \tag{23}$$

## B. Kinetic Energy

For a translational and vibrational motion shown in Figure 2(c), its kinetic energy of the diatomic molecule is calculated using two different ways. First, kinetic energy is determined from each atom based on equation (1) as shown in Figure 3(a). Second, kinetic energy is determined from translational and vibrational motion separately based on equation (2) as shown in Figure 3(b).

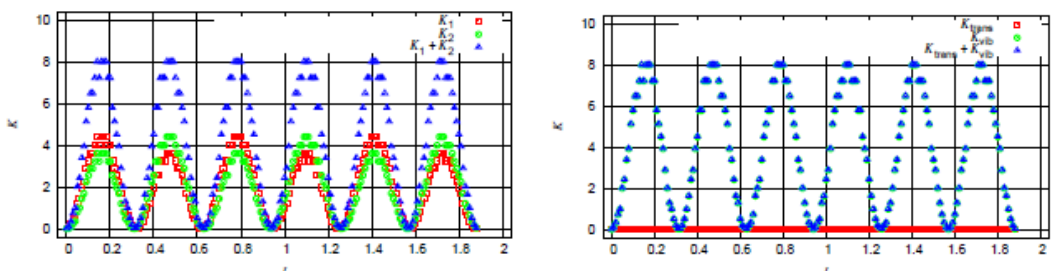

Figure 3. Kinetic energy-time graph for molecule motion in Figure 3(c):
(a) kinetic energy for each atom (b) kinetic energy for translational and vibrational motion separately.

Calculation of kinetic energy of an atom moving in translational-vibrational motion is not analytically simple, since it is not easy to express its mathematical function of its motion. However, it will be possible to determine its kinetic energy with the numerical method of molecular dynamic (Figure 3(a)). It can be seen from Figure 2(c) that its maximum magnitude of tangent (maximum speed) of $m_1$ in time interval 0 and 0.314 is at $t = 0.157$ and this maximum speed is slightly greater than that at $t = 0.471$ in time interval 0.314 and 0.628. It can also be observed that the maximum speed of $m_2$ is slightly greater than that of $m_1$ at time interval 0 and 0.314. Therefore, the maximum kinetic energy of $m_1$ is not exactly the same as $m_2$ and their values change alternately with time (Figure 3(a)).

Kinetic energy from translational motion and vibrational motion has been calculated in equations (15) and (18) respectively and its total is in equation (22). Since $K_{trans}$ is small compared to $K_{vib}$ then the total is nearly the same as $K_{vib}$ (Figure 3(b)).

## CONCLUSION

Motion of a diatomic molecule has been simulated using granular particles. For one-dimensional ideal gas, kinetic energy of a diatomic molecule can be calculated as a total of kinetic energy from each atom or a total of kinetic energy from translational motion and vibrational motion. Calculation of kinetic energy of each atom can be solved by numerical method of molecular dynamic simulation.

## ACKNOWLEDGMENTS

Authors would like to thank to Nuclear Physics and Biophysics Research Division, Faculty of Mathematics and Natural Sciences Insitut Teknologi Bandung for the friendly discussion atmosphere, and Hibah Kapasitas FMIPA.PN-6-18-2010 for partially support to this work.

## REFERENCES

1. Mark W Zemansky and Richard H Dittman, Heat and Thermodynamics, 7ed, McGraw-Hill Com Inc, Singapore, 1997, pp 323-324
2. Francis W Sears and Gerhard L Salinger, Thermodynamics, Kinetic Theory and Statistical Thermodynamics, 3ed, Addison-Wesley Pub com. 1986, Sydney, pp 370-372.
3. Daniel J Amit and Yosef Verbin, "Statistical Physics: an introductory course" World Scientific Pub com., 1999, pp 340-341.

4.  M. P. Allen and D. J. Tildesley, "Computer simulation of liquids", Oxford University Press, New York, 1989, pp. 82-85.

5.  Götz Giese and Annette Zippelius, "Collision Properties of One-Dimensional Granular Particles with Internal Degrees of Freedom", Physical Review E 45, 4828-4837 (1996)

6.  Ítalo'Ivo Lima Dias Pinto, Alexandre Rosas, and Katja Lindenberg, "Energy Transport in a One-Dimensional Granular Gas", Physical Review E 79, 061307 (2009)

7.  V. Yu. Zaburdaev, M. Brinkmann, and S. Herminghaus, "Free Cooling of the One-Dimensional We Granular Gas", Physical Review Letters 97, 018001 (2006)

8.  Rosa Ramirez and Patricio Cordero, "Kinetic Description of a Fluidized One-Dimensional Granular System", Physical Review E 59, 656-654 (1999)